# Integrated Circuits Based on Bilayer MoS$_2$ Transistors

H. Wang[1,*,†], L. Yu[1,†], Y.-H. Lee[1,3], Y. Shi[1], A. Hsu[1], M. Chin[2], L.-J. Li[3], M. Dubey[2], J. Kong[1], T. Palacios[1,*]

[1]Department of Electrical Engineering and Computer Science, Massachusetts Institute of Technology, 77 Massachusetts Avenue, Cambridge MA 02139, USA. Tel: +1 (617) 324-2395.

[2]United States Army Research Laboratory, 2800 Powder Mill Road, Adelphi, MD 20783-1197, USA.

[3]Institute of Atomic and Molecular Sciences, Academia Sinica, Taipei, 11529, Taiwan.

*Corresponding author E-mail: hanw@mtl.mit.edu, tpalacios@mit.edu.

[†]H. W. and L. Y. contributed equally to this work.

**Abstract**

Two-dimensional (2D) materials, such as molybdenum disulfide (MoS$_2$), have been shown to exhibit excellent electrical and optical properties. The semiconducting nature of MoS$_2$ allows it to overcome the shortcomings of zero-bandgap graphene, while still sharing many of graphene's advantages for electronic and optoelectronic applications. Discrete electronic and optoelectronic components, such as field-effect transistors, sensors and photodetectors made from few-layer MoS$_2$ show promising performance as potential substitute of Si in conventional electronics and of organic and amorphous Si semiconductors in ubiquitous systems and display applications. An important next step is the fabrication of fully integrated multi-stage circuits and logic building blocks on MoS$_2$ to demonstrate its capability for complex digital logic and high-frequency ac applications. This paper demonstrates an inverter, a NAND gate, a static random access memory, and a five-stage ring oscillator based on a direct-coupled transistor logic technology. The circuits comprise between two to twelve transistors seamlessly integrated side-by-side on a single sheet of bilayer MoS$_2$. Both enhancement-mode and depletion-mode transistors were fabricated thanks to the use of gate metals with different work functions.

Two-dimensional (2D) materials, such as molybdenum disulfide (MoS$_2$)[1] and other members of the transition metal dichalcogenides family, represents the ultimate scaling of material dimension in the vertical direction. Nano-electronic devices built on 2D materials offer many benefits for further miniaturization beyond Moore's Law[2,3] and as a high-mobility option in the emerging field of large-area and low-cost electronics that is currently dominated by low-mobility amorphous silicon[4] and organic semiconductors[5,6]. MoS$_2$, a 2D semiconductor material, is also attractive as a potential complement to graphene[7,8,9] for constructing digital circuits on flexible and transparent substrates, while its 1.8 eV bandgap[10,11] is advantageous over silicon for suppressing the source-to-drain tunneling at the scaling limit of transistors[12]. Recently, various basic electronic components have been demonstrated based on few-layer MoS$_2$, such as field-effect transistors (FETs)[13,14,15], sensors[16] and phototransistors[17]. However, until now, only primitive circuits involving one or two discrete MoS$_2$ transistors connected through external wiring have been reported[18]. These devices also have mismatched input and output logic levels, making them unsuitable for cascading multiple logic stages. This paper addresses the next challenge in the development of 2D nanoelectronics and optoelectronics on MoS$_2$, that is the construction of fully integrated multi-stage logic circuits based on this material to demonstrate its capability for complex digital logic. These circuits were fabricated entirely on the same chip for the first time thanks to the seamless integration of both depletion-mode (D-mode) and enhancement-mode (E-mode) MoS$_2$ transistors. The transistors show multiple state-of-the-art characteristics, such as current saturation, high on/off ratio (>10$^7$), and record on-state current density (>23 µA/µm). This demonstration of integrated logic gates, memory elements and a ring oscillator operating at 1.6 MHz represents an important step towards developing 2D electronics for both conventional and ubiquitous applications, offering materials that can combine silicon-like performance with the mechanical flexibility and integration versatility of organic semiconductors.

Molybdenum disulfide (MoS$_2$) is a layered semiconductor from the transition metal dichalcogenides material family (TMD), MX$_2$ (M=Mo, W; X=S, Se, Te)[10,11,19,20]. A single molecular layer of MoS$_2$ consists of a layer of Mo atoms sandwiched between two layers of sulfur atoms by covalent bonds[10]. The strong intra-layer covalent bonds confer MoS$_2$ crystals excellent mechanical strength, thermal stability up to 1090 °C in inert environment[21], and a surface free of dangling bonds. On the other hand, the weak inter-layer Van der Waal's force allows single- or few-layer MoS$_2$ thin films to be created through micro-mechanical cleavage technique[22] and through anisotropic 2D growth by chemical vapor deposition[23,24]. This unique property of $MoS_2$, and 2D materials in general, enables the creation of atomically smooth material sheets and the precise control on its number of molecular layers. Field-effect transistors (FETs) built on the ultra-thin few-layer 2D crystals, hence, are effectively the optimal form of ultra-thin body FETs[25], a transistor structure ideal for suppressing the short-channel effects at its scaling limit. This benefit of 2D FETs has been demonstrated in ref. [14], which shows a high on/off current ratio of $10^8$ in a single-layer $MoS_2$ FET.

The planar nature and mechanical flexibility of 2D materials also make them excellent candidates for fabricating light-weight and rollable or foldable electronic systems on common commodities like paper, plastics and textiles; as well as for constructing low-cost driving circuits for flat-panel display applications[26]. The incumbent technology for such applications is based on amorphous Si and polycrystalline Si, with organic semiconductors being the other potential option. Thin film transistors based on amorphous Si, for example, often have mobility below 1 $cm^2 V^{-1} s^{-1}$, the on-off ratio in excess of $10^6$, and the device switches from on to off within about 5–8 V, making it barely fit within the requirements of display applications[4]. The organic semiconductors offer ease of fabrication, but exhibit similar or even lower mobility than amorphous Si due to its intrinsically disordered nature and quantum-mechanical-tunneling based transport[27]. In contrast, the covalently-bonded, highly-ordered crystalline 2D materials have carrier mobility that is orders of magnitude higher than in amorphous Si and organic semiconductors. 2D materials are thus promising for improving the performance and enable new functionality of ubiquitous electronics and display technology, such as flexible radio-frequency identification tags and enhanced integration of drivers and logic circuits into display backplanes.

In this letter, we address the next key challenge in the development of 2D nanoelectronics by demonstrating the first fully integrated multi-stage circuits entirely assembled on few-layer $MoS_2$. These circuits are based on the development of a direct-coupled FET logic (DCFL)[28] in this material system, for which both enhancement-mode and depletion-mode devices with excellent pinch-off and current saturation are necessary. All the circuits were fabricated on bilayer $MoS_2$ obtained from micro-mechanical cleavage. Bilayer $MoS_2$ offers an excellent trade-off between the off-state and on-state current levels (see Supplementary Information). The number of $MoS_2$ layers can be confirmed by atomic force microscopy (AFM) (Fig. 1A) based on its thickness and by Raman spectroscopy based on the peak spacing between the $E_{2g}$ mode and the $A_{1g}$ mode[29], respectively (Fig. 1B and Supplementary

Information Fig. S1). The lateral size of the exfoliated bilayer MoS$_2$ thin films, which are 13 Å thick, can reach up to 40 μm (Inset of Fig. 1A).

The direct-coupled FET logic (DCFL) technology is a popular architecture for constructing high-speed circuit with low power dissipation, where an excellent trade-off between speed and power loss may be achieved[30,31] and is suitable for application in low-power flexible electronics. The DCFL circuits used in this letter integrates both negative (D-mode) and positive (E-mode) threshold voltage transistors on the same chip (Fig. 2A and 2B). This can be achieved through engineering the gate metal work functions of the MoS$_2$ FETs (see Supplementary Information). Fig. 2C and 2D shows the device characteristics of two MoS$_2$ FETs with Al ($w_M$ = 4.08 eV) and Pd ($w_M$ = 5.12-5.60 eV)[32] gates, respectively, fabricated side-by-side on the same bilayer MoS$_2$ thin film (see Supplementary Information for fabrication). The difference in the work functions of these two metals effectively shifts the threshold voltages of the MoS$_2$ FET characteristics by about 0.76 V to form a D/E-FET pair (Fig. 2C and 2D). The shift in the threshold voltage is lower than the metal work function difference in vacuum (~1.04 eV) (see Supplementary Information).

Both the D-mode and E-mode FETs have a high on/off current ratio in excess of $10^7$ (Fig. 2D), which is very close to that in single-layer MoS$_2$ FETs[14]. On the other hand, at the on-state, these devices based on bilayer MoS$_2$ have much higher on-state current density (exceeding 23 μA/μm at $V_{ds}$=1.0 V and $V_{tg}$= 2.0 V for the depletion mode FET, Fig. 2C) than that reported for single-layer MoS$_2$ FETs[14,18]. The corresponding maximum transconductance of the bilayer FETs exceeds 12 μS/μm at $V_{ds}$=1.0 V. These bilayer MoS$_2$ FETs can hence offer superior on-state performance than single-layer devices, with only a small degradation in terms of on/off current ratio. The high-field transport of both FETs shows saturation behavior (Fig. 3A), a critical feature for both logic and analog circuits, for the first time in top-gate MoS$_2$ FETs. The excellent match between the on-set of saturation and the gate overdrive (i.e. $V_{sat}=V_{tg}-V_t$, where $V_{sat}$ is the saturation voltage and $V_t$ is the threshold voltage of the FETs) indicates that the current saturation is due to the classic channel pinch-off mechanism, as is typical for long channel MOSFETs[33]. The field-effect mobility at $V_{ds}$=1 V is extracted to be 10-15 cm$^2$ V$^{-1}$ s$^{-1}$ before depositing the halfnium oxide (HfO$_2$). After the MoS$_2$ thin film was passivated by a 20-nm-thick HfO$_2$ layer, a conservative estimate based on back-gated characteristics shows a field effect mobility exceeding 300 cm$^2$ V$^{-1}$ s$^{-1}$ (see Supplementary Information).

Using the technology described above, we built four different integrated logic circuits entirely assembled on bilayer $MoS_2$: a logic inverter, a NAND gate, a static random access memory (SRAM) cell, and a 5-stage ring oscillator, all constructed with DCFL technology[28]. For each of the four logic circuits, all active and passive elements are integrated on the same piece of bilayer $MoS_2$. It is found that a supply voltage of $V_{dd}$=2 V is suitable for operating the fabricated circuits. Hence, in this letter, a voltage level close to 2 V represents the logic state 1 while a voltage level close to 0 V represents the logic state 0.

An inverter circuit is a basic logic element that outputs a voltage representing the opposite logic-level to its input. Our inverter was constructed from an enhancement-mode $MoS_2$ transistor, and a depletion-mode resistor that was formed by connecting the gate of a depletion-mode transistor directly to its source electrode (Fig. 3B inset and Fig. 5A). The quality of a logic inverter is often evaluated using its voltage transfer curve (Fig. 3B), which is a plot of input vs. output voltage. When the input voltage is $V_{in}$=2 V (logic state 1), the E-mode $MoS_2$ FET is much more conductive than the depletion-mode FET, setting the output voltage to below 0.2 V (logic state 0). When $V_{in}$ is 0 V (logic state 0), the $MoS_2$ FET is non-conducting and the output is close to 2 V (logic state 1). The slope of the transition region in the middle provides a measure of the gain - or the quality of switching. In the circuit of Fig. 3B, a voltage gain close to 5 is achieved. Fig. 3B also shows the mirror reflection of the $V_{in}$-$V_{out}$ characteristics, which highlights the robustness of the inverter towards noise for multi-stage operations. When multiple inverter stages are cascaded together, the output signal from the previous stage becomes the input signal to the next stage. Hence, the shaded area ($NM_L$ and $NM_H$) represents the noise margin that can be tolerated by the inverter for multi-stage operations, which is particularly important for the demonstration of the ring oscillator.

The schematic design and the optical micrograph of an NAND gate circuit fabricated on a sheet of bilayer $MoS_2$ are shown in Fig 4A. The output of the circuit is close to 2 V (logic state 1) when either or both of the inputs are at logic state 0 ($V_{in}$ <0.5 V). Under this state, at least one of the $MoS_2$ FETs is non-conducting and the output voltage is clamped to the supply voltage $V_{dd}$. The output is at logic state 0 only when both inputs are at logic state 1, so that both $MoS_2$ FETs are conducting. In Fig. 4C, the output voltage is measured as a function of time while the two input voltage states vary across all four possible logic combinations (0,0), (0,1), (1,0), and (1,1). This data demonstrates the stable NAND gate functions of this two-transistor bilayer $MoS_2$ circuit. A NAND gate is one of the two basic logic gates (the other being NOR gate) with universal functionality. Any other type of logic gates (AND, OR, NOR,

XOR, etc.) can then be constructed with a combination of NAND gates. Hence, this first demonstration of a NAND gate shows that it is possible to fabricate any kind of digital integrated circuit with $MoS_2$ thin film layers.

A flip-flop memory element (SRAM) has also been constructed from a pair of cross-coupled inverters (Fig. 4A). This storage cell has two stable states at the output, which are denoted as 0 and 1. The flip-flop cell can be set to logic state 1 (or 0) by applying a low (or high) voltage to the input. To verify the functionality of this flip-flop cell, a voltage source is applied to the input to set $V_{in}$ to 2 V at time T=0 s. This drives $V_{out}$ into logic state 0 (Fig. 4B). Then at T=20 s, the switch at $V_{in}$ is opened and the output of the SRAM cell $V_{out}$ remains at logic state 0. At time T=60 s, we apply $V_{in}$=0 V at the input to write a logic state 1 into $V_{out}$. As the switch is opened again at T=80 s, the output of the SRAM cell remains in the logic state 1. This data demonstrates that the flip-flop SRAM circuit fabricated on the bilayer $MoS_2$ thin film indeed functions as a stable memory cell.

Finally, a 5-stage ring oscillator was constructed to assess the high frequency switching capability of $MoS_2$ and for evaluating the material's ultimate compatibility with conventional circuit architecture[28,34,35,36] (Fig. 5A). The ring oscillator, which integrates 12 bilayer $MoS_2$ FETs together, was realized by cascading five inverter stages in a close loop chain (Fig. 5B). An extra inverter stage was used to synthesize the output signal by isolating the oscillator operation from the measurement setup to prevent the interference between them. The output of the circuit was connected to either an oscilloscope or a spectrum analyzer for evaluation. The voltage transfer curve of the test inverter circuit fabricated side-by-side on the same piece of bilayer $MoS_2$ thin film (Fig. 3B and Fig. 5A) as the ring oscillator, shows that the gain in each inverter stage is close to 5. For robust ring oscillator performance, it is imperative to have stable operations in all five inverter stages throughout the oscillation cycles, and its tolerance towards noise can be determined from the noise margins for both low and high logic levels, i.e. the shaded regions in Fig. 3B. The positive feedback loop in the ring oscillator results in a statically unstable system, and the voltage at the output of each inverter stage oscillates as a function of time (Fig. 5C). At $V_{dd}$=2 V, the fundamental oscillation frequency is at 1.6 MHz, corresponding to a propagation delay of $\tau_{pd} = 1/(2nf) = 62.5$ ns per stage, where *n* is the number of stages and *f* is the fundamental oscillation frequency. The frequency performance of this ring oscillator, while operating at a much lower $V_{dd,}$ is at least an order of magnitude better than the fastest integrated organic semiconductor ring oscillators[35]. It also rivals the speed of ring oscillators constructed from the printed ribbons of single-crystalline silicon reported in the literature[37]. The output voltage swing measured by the oscilloscope (input impedance 1 MΩ) is about 1.2 V.

The output signal of the ring oscillator can also be measured in terms of its frequency power spectrum. Fig. 5D shows the spectrum of the output signal from the ring oscillator as a function of the drain bias voltage $V_{dd}$ (Fig. 5D). The resonance frequency is at 0.52 MHz for $V_{dd}$=1.15 V. The corresponding fundamental resonance frequency reaches 1.6 MHz as $V_{dd}$ increases to 2 V. The improvement in frequency performance with increasing $V_{dd}$ can be attributed to the enhancement in the current driving capability of the ring oscillator due to the rise in the drain current $I_{ds}$ in each individual $MoS_2$ FET with increasing drain and gate voltages. The fundamental frequency of oscillation is currently limited by the parasitic capacitances in various parts of the circuit rather than the intrinsic performance of the $MoS_2$ devices (see Supplementary Information). The signal peaks measured by the spectrum analyzer increases from −65 dBm to -46 dBm as $V_{dd}$ raises from 1.15 V to 2V. This is again a result of the $I_{ds}$ dependence on $V_{dd}$.

To summarize, the realization of fully integrated multi-stage logic circuits based on few-layer $MoS_2$ DCFL represents the first demonstration of integrated multi-stage systems on any 2D materials, including graphene. It is an important step towards realizing 2D nanoelectronics for high performance low-power applications. Further optimization is underway to increase operating speed, and towards realizing complementary logic circuits to decrease the power dissipation. With the rapid progress in large-scale growth of $MoS_2$ by chemical vapor deposition[23,24], these 2D crystals are extremely promising new materials for both conventional and ubiquitous electronics.

**Acknowledgement**

The authors are grateful to D. Antoniadis and L. Wei for discussions. The authors acknowledge financial support from the Office of Naval Research (ONR) Young Investigator Program, the Microelectronics Advanced Research Corporation Focus Center for Materials, Structure and Device (MARCO MSD), National Science Foundation (NSF DMR 0845358) and the Army Research Laboratory. This research has made use of the MIT MTL and Harvard CNS facilities.

**Additional information**

The authors declare no competing financial interests.

**Figures**

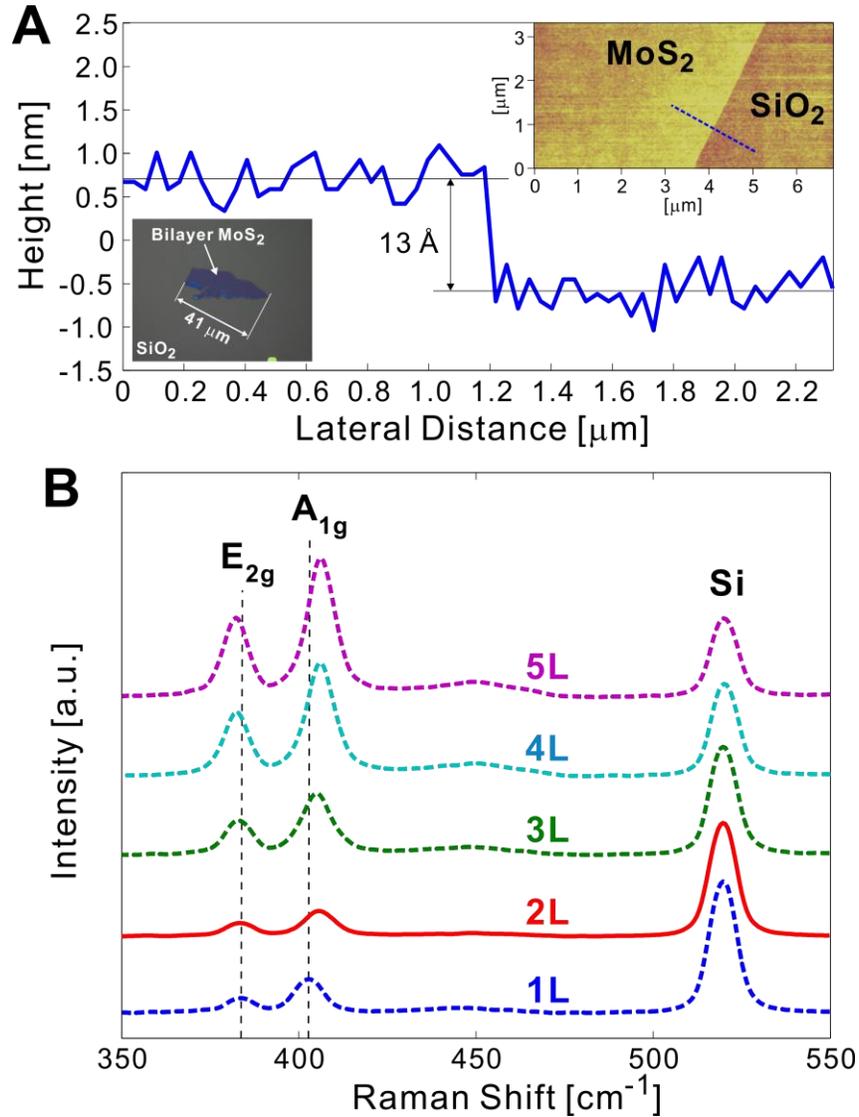

**Fig. 1.** Optical micrograph, AFM and Raman spectroscopy of bilayer $MoS_2$. **(A)** Optical micrograph and AFM data of a bilayer $MoS_2$ thin film. The flake is 13 Å thick, which is equal to twice the thickness of single-layer $MoS_2$, confirming the flake being bilayer. **(B)** The number of layers in the $MoS_2$ thin film can also be confirmed from its Raman spectroscopy based on the peak spacing between the $E_{2g}$ mode and the $A_{1g}$ mode[29]. The red-shift of $E_{2g}$ peak and blue-shift of $A_{1g}$ peak lead to increasing peak spacing between $E_{2g}$ and $A_{1g}$ modes as the number of layers in the $MoS_2$ thin film increases.

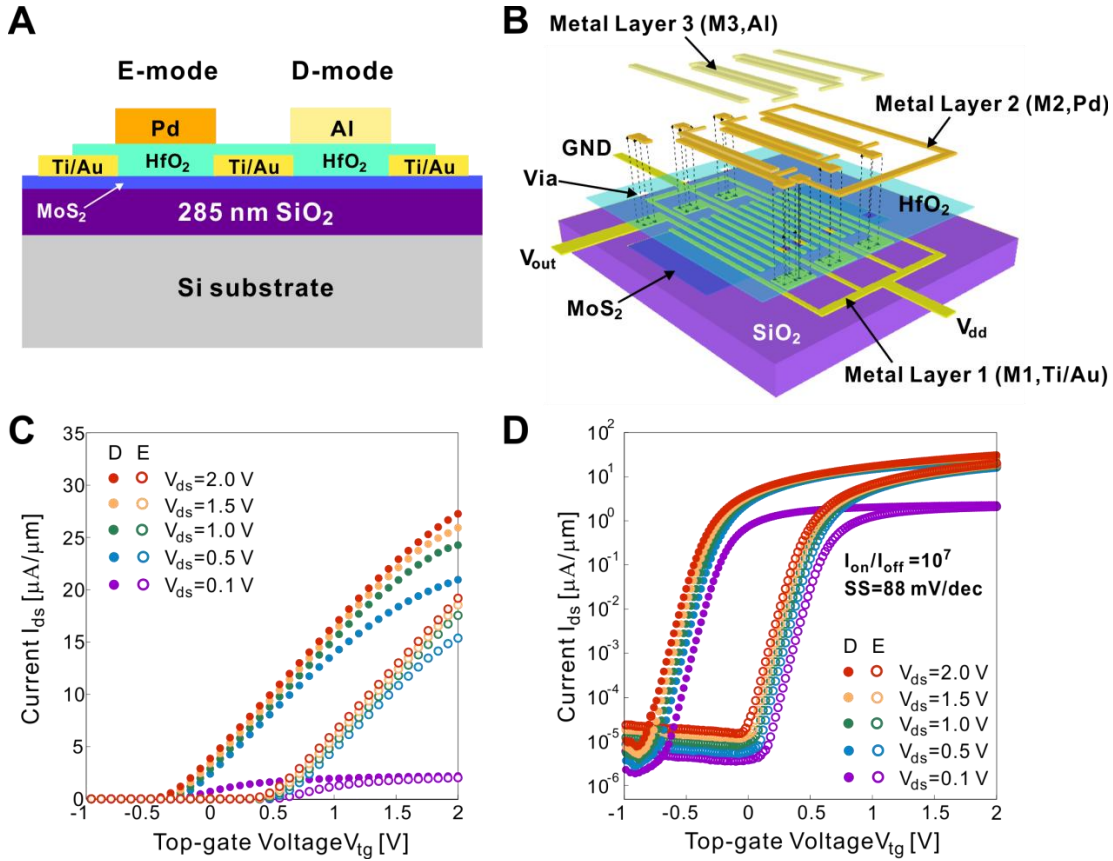

**Fig. 2. (A)** Schematic representation of an E-mode and a D-mode device. **(B)** Schematic illustration of an integrated 5-stage ring oscillator circuit on $MoS_2$ thin films, which is constructed by integrating 12 $MoS_2$ FETs. Three distinct metal layers of the $MoS_2$ IC are represented by M1, M2, and M3. M1 is directly in contact with the bilayer $MoS_2$ thin film while M2 and M3 are the Pd and Al gate layers, respectively. Via holes are etched through the $HfO_2$ dielectric layer to allow connections from M2 and M3 to M1. The fabricated ring oscillator circuit corresponding to the design above is shown in Fig. 5. The general aspects of the fabrication process apply to all the devices and logic circuits presented in this letter. **(C)** The transfer characteristics of depletion (D) mode and enhancement (E) mode bilayer $MoS_2$ FETs. The depletion mode FET has Al as the gate metal while the enhancement FET has Pd as the gate metal. The on-state current and transconductance of a device are its key dc performance metrics, critical for circuit application. In these bilayer $MoS_2$ FETs, the on-state current density exceeds 23 μA/μm at $V_{ds}$=1 V and the transconductance is above 12 μS/μm, both being the highest values reported for $MoS_2$ FETs so far. The difference between the work functions of Al and Pd (~1.04 V in vacuum) gates results in a 0.76 V shift in the threshold voltage. The discrepancy between the work function difference in vacuum and in $HfO_2$ can be attributed to the dipoles at the metal/$HfO_2$ interface, resulting from charge transfer across this boundary[38]. **(D)** The transfer characteristics in logarithmic scale of depletion (D) mode and enhancement (E) mode bilayer $MoS_2$ FETs. The $I_{on}/I_{off}$ ratio exceeds $10^7$ for $V_{ds}$ above 0.5 V, and is about $10^6$ at $V_{ds}$=0.1 V. The sub-threshold slope (SS) is 88 mV/dec. Device dimension: $L_g$=1 μm and $L_{ds}$=1 μm. The substrate is grounded.

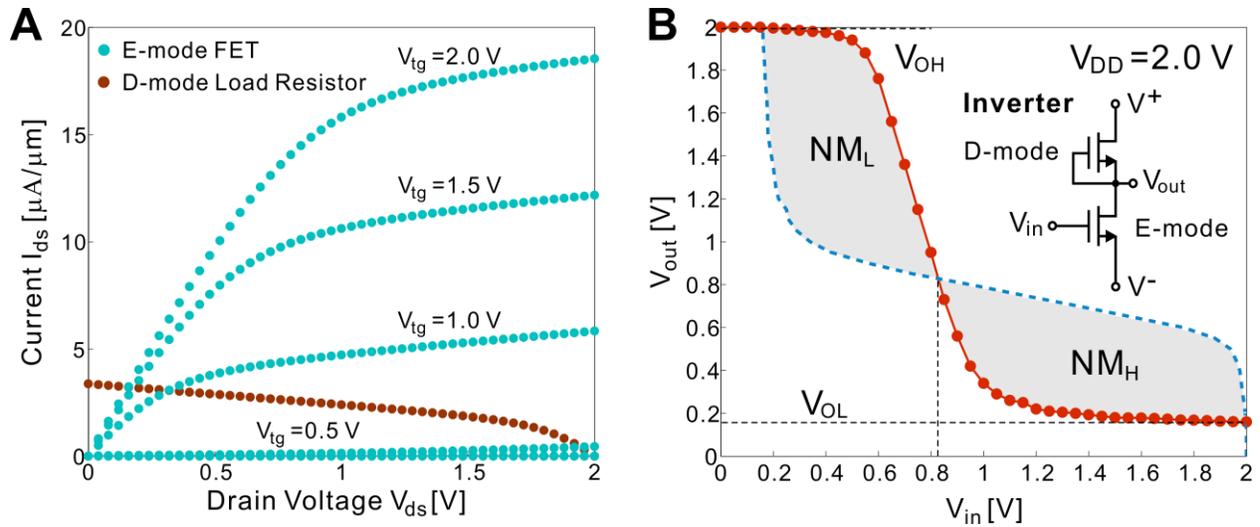

**Fig. 3.** Demonstration of an integrated logic inverter on bilayer $MoS_2$. **(A)** Output characteristics ($I_{ds}$-$V_{ds}$) of the E-mode FET and D-mode load for the inverter shown in Fig. 5A. $L_g$=1 μm and $L_{ds}$=1 μm. For the E-mode FET, in the linear regime at small source-drain voltages, the current is proportional to $V_{ds}$, indicating that the source and drain electrodes made of Ti/Au metal stack forms ohmic contact with $MoS_2$. The current saturates at higher drain bias ($V_{ds}$>$V_g$-$V_t$) due to the formation of depletion region on the drain side of the gate, as is typical of long channel MOSFETs. **(B)** Output voltage as a function of the input voltage, and its mirror reflection, for a bilayer $MoS_2$ logic inverter. The shaded area indicates its noise margins ($NM_L$ and $NM_H$) for logic operation. The gain of the inverter is close to 5. (Inset) Schematic of the electronic circuit for a logic inverter.

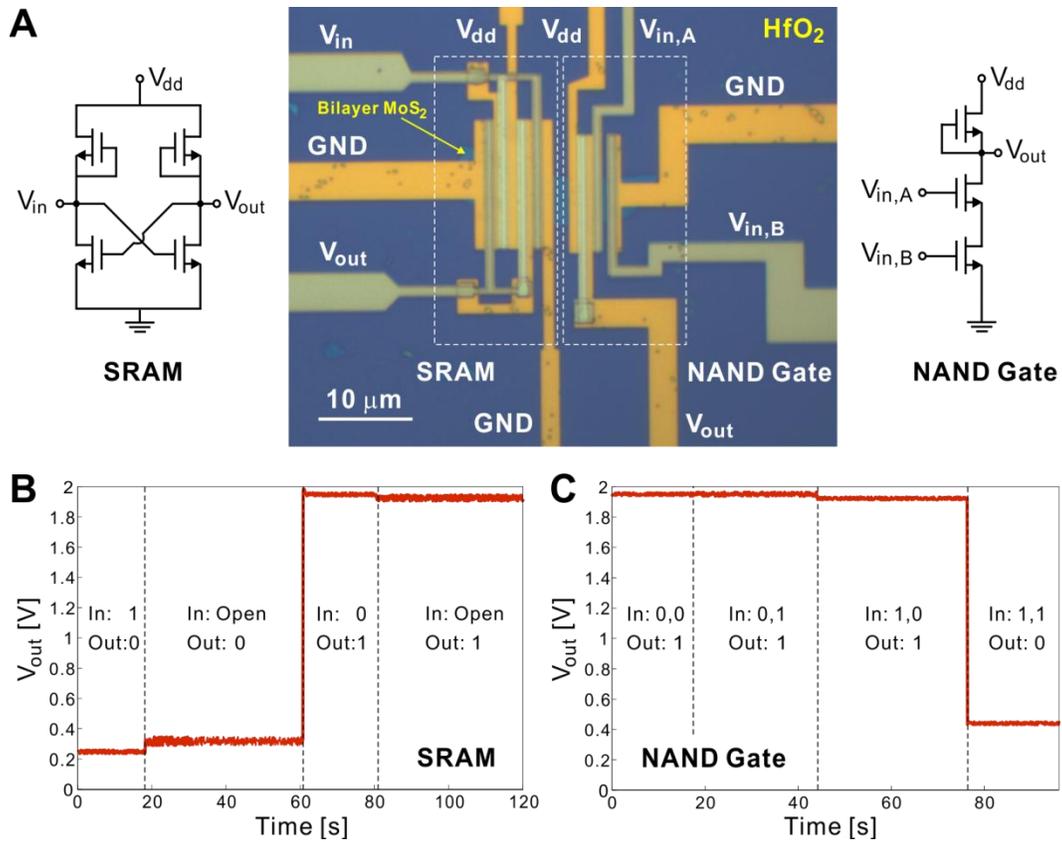

**Fig. 4.** Demonstration of an integrated NAND logic gate and a static random-access memory (SRAM) cell on bilayer $MoS_2$. **(A)** Optical micrograph of the NAND gate and the SRAM fabricated on the same bilayer $MoS_2$ thin film. The corresponding schematics of the electronic circuits for the NAND gate and SRAM are also shown. **(B)** Output voltage of the flip-flop memory cell (SRAM). A logic state 1 (or 0) at the input voltage can set the output voltage to logic state 0 (or 1). In addition, the output logic state stays at 0 or 1 after the switch to the input has been opened. **(C)** Output voltage of the NAND gate for four different input states: (0,0), (0,1), (1,0), and (1,1). A low voltage below 0.5 V represents a logic state 0 and a voltage close to 2 V represents a logic state 1.

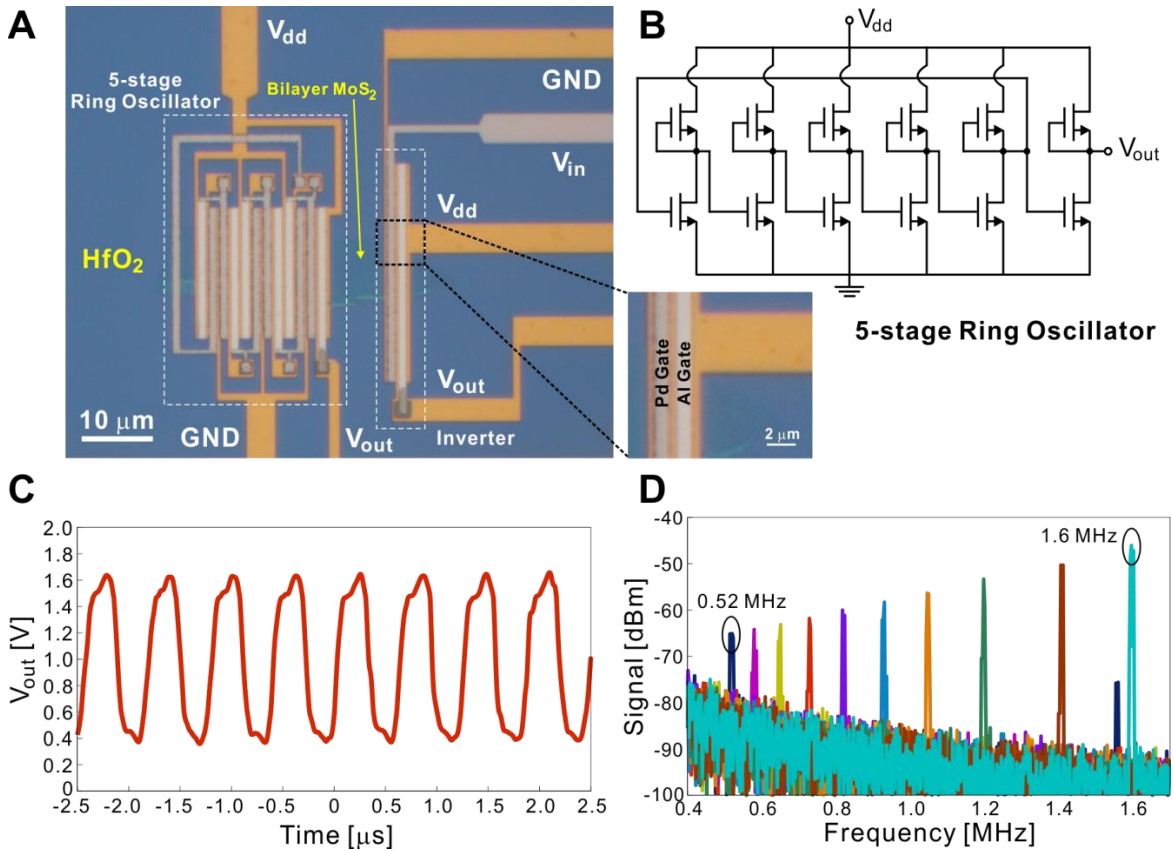

**Fig. 5.** A 5-stage ring oscillator based on bilayer $MoS_2$. **(A)** Optical micrograph of the ring oscillator constructed on a bilayer $MoS_2$ thin film. **(B)** Schematic of the electronic circuit of the 5-stage ring oscillator. The first five inverter stages form the positive feedback loop, which leads to the oscillation in the circuit. The last inverter serves as the synthesis stage. **(C)** Output voltage as a function of time for the ring oscillator at $V_{dd}$=2 V. The fundamental oscillation frequency is at 1.6 MHz. The corresponding propagation delay per stage is 62.5 ns. **(D)** The power spectrum of the output signal as a function of $V_{dd}$. From left to right, $V_{dd}$= 1.15 V, and 1.2 to 2.0 V in step of 0.1 V. The corresponding fundamental oscillation frequency increases from 0.52 MHz to 1.6 MHz.

**Synopsis for Table of Contents**

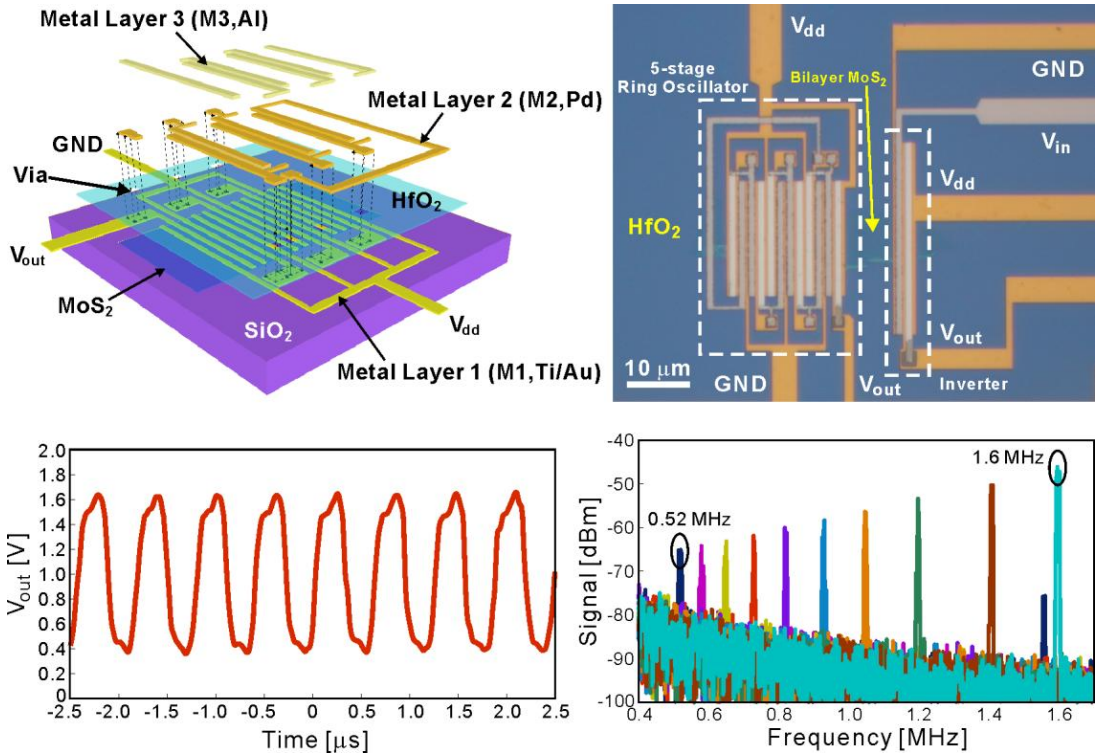

# Supplementary Information

## Integrated Circuits Based on Bilayer MoS$_2$ Transistors

H. Wang$^{1,*,†}$, L. Yu$^{1,†}$, Y.-H. Lee$^{1,3}$, Y. Shi$^1$, A. Hsu$^1$, M. Chin$^2$, L.-J. Li$^3$, M. Dubey$^2$, J. Kong$^1$, T. Palacios$^{1,*}$

$^1$Department of Electrical Engineering and Computer Science, Massachusetts Institute of Technology, 77 Massachusetts Avenue, Cambridge MA 02139, USA. Tel: +1 (617) 324-2395.

$^2$United States Army Research Laboratory, 2800 Powder Mill Road, Adelphi, MD 20783-1197, USA.

$^3$Institute of Atomic and Molecular Sciences, Academia Sinica, Taipei, 11529, Taiwan.

*Corresponding author E-mail: hanw@mtl.mit.edu, tpalacios@mit.edu.

$^†$H. W. and L. Y. contributed equally to this work.

**Methods**

***Device and integrated circuit fabrication:*** the fabrication of our devices and circuits starts with the exfoliation of MoS$_2$ thin films from commercially available bulk MoS$_2$ crystals (SPI Supplies) onto 285 nm SiO$_2$ on Si substrate, which has pre-patterned alignment grids (Cr/Au), using the micro-mechanical cleavage technique. The thickness of the SiO$_2$ was selected to provide the optimal optical contrast for locating MoS$_2$ flakes relative to the alignment grids and for identifying their number of layers[S1]. The number of MoS$_2$ layers was then confirmed by atomic force microscopy (AFM) based on its thickness and by Raman spectroscopy based on the peak spacing between the $E_{2g}$ mode and the $A_{1g}$ mode, respectively[S2]. The sample was then annealed at 350 C° in Ar 600 sccm/H$_2$ 30 sccm for three hours to clean away the tape residue. The next step was to pattern the first metal layer (M1), which are the electrodes directly in contact with MoS$_2$, i.e. source and drain of the devices, using electron-beam lithography (Elionix F125) based on poly(methyl methacrylate) (950k MW PMMA) resist. We then evaporated 3 nm Ti/ 50 nm Au followed by lift-off to form the contacts. Subsequently, the samples were annealed again at 350 ℃, 600 sccm Ar/30 sccm H$_2$ for three hours. This annealing step reduces device resistance and also removes the PMMA residue to create a clean surface for subsequent atomic layer deposition (ALD) process. The top gate dielectric consisting of 20 nm HfO$_2$ was then deposited by ALD. To fabricate discrete transistors, the last step of the fabrication was to pattern the top gate electrode by electron-beam lithography, which was then formed by depositing the desired gate metal. The devices reported in this work (Fig. 2 in the main text) have L$_G$=1 μm and L$_{DS}$=1 μm. The gate is aligned to the channel using the standard alignment techniques in e-beam lithography. For the construction of integrated logic circuits (Fig. 2B in the main text), the second and third metal layers (M2 and M3) need to be connected to the first metal layer (M1) at certain locations depending on the design. This was achieved by patterning and etching via holes through the HfO$_2$ dielectric using reactive ion etching (RIE) with BCl$_3$/Cl$_2$ gas chemistry. This etching step preceded the definition of the gate metal layers M2 and M3.

***AFM and Raman spectroscopy:*** Atomic force microscopy (AFM) for identifying the thin film thickness was performed on a Veeco Dimension$^{\text{TM}}$ 3100 system. Raman spectroscopy was performed with a 532 nm Nd:YAG laser. All optical micrographs were taken with a Zeiss Axio Imager.A1m microscope.

***ALD and Via Hole Etching:*** the HfO$_2$ gate dielectric was deposited using ALD at 170 °C. The ALD deposition of HfO$_2$ was done on a commercial Savannah ALD system from Cambridge NanoTech using alternating cycles of H$_2$O and tetrakis(dimethylamido)hafnium (TDMAH) as the precursors. To fabricate the integrated circuits shown in this report, it is also necessary to etch via holes through the HfO$_2$ dielectric so that on-chip interconnections can be made between metal layer 1 and metal layers 2 and 3. We used a commercial Electron Cyclotron Resonance Reactive Ion Etcher (ECR/RIE) system (Plasma Quest) to perform this etch using BCl$_3$/Cl$_2$ gas chemistry. The ratio between the flow rates of BCl$_3$ and Cl$_2$ is 4:1. The etch rate of our low power recipe is around 6 nm/min.

***Device and circuit characterization:*** Device characterization was performed using an Agilent 4155C semiconductor parameter analyzer and a Lakeshore cryogenic probe station with micromanipulation probes. The integrated circuits were characterized with an Agilent 54642A oscilloscope (1 MΩ input impedance) and the output signal power spectrum of the ring oscillator was measured with an Agilent N9010A Signal Analyzer (50 Ω input impedance). All measurements were done in vacuum (~10$^{-5}$ Torr) at room temperature.

## Identifying the number of layers in exfoliated MoS$_2$ thin film by AFM and Raman

The number of molecular layers in exfoliated MoS$_2$ thin film can be identified from measuring its thickness by atomic force microscope (AFM) and by Raman spectroscopy, based on the peak spacing between the $E_{2g}$ mode and $A_{1g}$ mode[S2]. Fig. S1 shows the correspondence between the AFM data and Raman data for 1-layer (1L) to 5-layer (5L) MoS$_2$ flakes. After exfoliating the MoS$_2$ thin film onto 285 nm SiO$_2$/Si substrate, the flakes were first located using optical microscope and the number of molecular layers was estimated based on their optical contrast[S1]. Then the number of layers was confirmed using both AFM and Raman spectroscopy. Bilayer MoS$_2$ thin films used in this report typically have a thickness around 13 Å as measured by AFM.

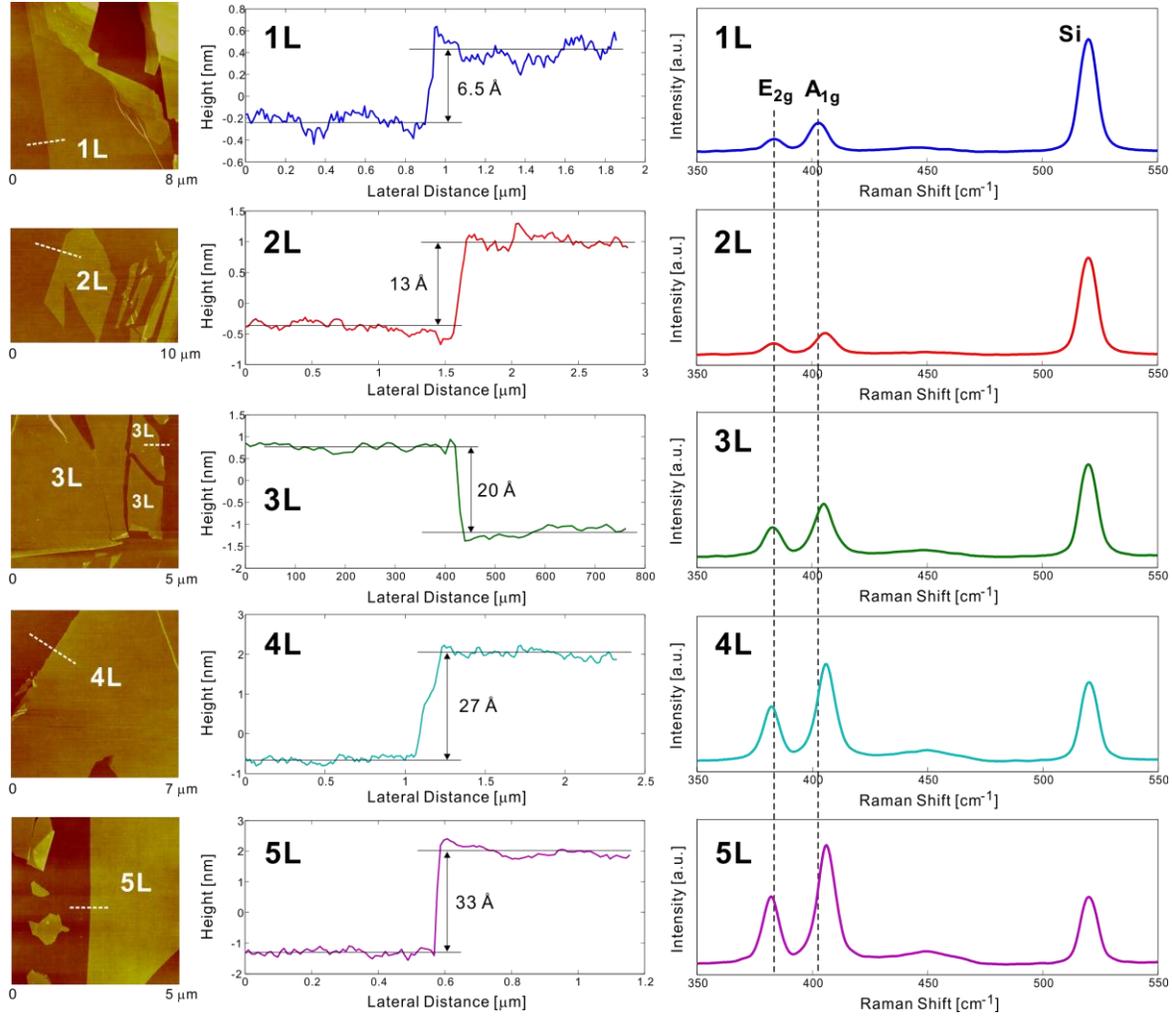

**Fig. S1.** Corresponding data from AFM and Raman spectroscopy measurements for one-layer to five-layer $MoS_2$ thin films. For each $MoS_2$ flake, the height information from AFM and Raman peak spacing between $E_{2g}$ mode and $A_{1g}$ mode are cross-checked to confirm the number of layers in the $MoS_2$ thin film.

Although single-layer $MoS_2$ FETs exhibits high on/off current ratio and low off-state current, which is important for minimizing the loss in the devices when they are turned off, it can only supply a very limited amount of current when the device is turned on (only 2.5 µA/µm at $V_{ds}$=0.5 V as reported in ref. S3). Since the speed of a logic circuit is often determined by the ratio between the charge required to change the voltage across the various capacitances in the circuits and the current that can be supported by the transistors, the low on-state current in single-layer $MoS_2$ may limit the operation speed of any electronic systems constructed from this material. On the other hand, by increasing the number of $MoS_2$ layers, the on-state current of $MoS_2$ FETs can be increased significantly (close to 20 µA/µm at $V_{ds}$=0.5 V $V_{tg}$=2 V for bilayer $MoS_2$ in this work) with only small degradation in terms of on/off current ratio. The higher on-state current in bilayer $MoS_2$ as compared to single layer $MoS_2$ can be due to several factors, with the most important one being the increase in mobility with the number of $MoS_2$ molecular layers. The mobility of $MoS_2$ increases with the number of molecular layers, as reported in ref. S4. This is also evidenced by other recent results in the literature. For example, ref. S5 reported a mobility of 517 $cm^2$ $V^{-1}$ $s^{-1}$ for a 23 layer flake (15 nm thick). In this work, the bi-layer $MoS_2$ flake shows a mobility of 313 $cm^2$ $V^{-1}$ $s^{-1}$ while the best mobility reported for single layer $MoS_2$ flake is 217 $cm^2$ $V^{-1}$ $s^{-1}$ [S3]. For these reasons, we select bi-layer $MoS_2$ thin film as the material on which we demonstrate integrated logic circuits. For real electronic applications in the future, the selection of the number of layers may depend on the type of application. If better frequency performance is needed, then multi-layer $MoS_2$ may be used. If ultra-low power performance is necessary, then single-layer $MoS_2$ may be a better choice. And bi-layer and tri-layer $MoS_2$ thin films may offer good trade-off in between. In short, the capability to control the number of molecular layers in the 2D crystal and the consequent control of the electronic properties enables added flexibility in this material system. In the future, it may be possible to build integrated circuits where different sections of the IC use different number of layers of $MoS_2$ thin films. The high performance sections (e.g. analog to digital converters, high speed oscillators) can use multilayer $MoS_2$ thin films while the low loss section (e.g. memory units) can use fewer layer of $MoS_2$ thin films.

**Work function Difference between Al and Pd**

To build both depletion-mode and enhancement-mode FETs on the same sheet of $MoS_2$, we use metals with different work functions, $w_M$, as the gates to control the threshold voltages of the FETs. Fig. S2 shows the band diagram of FETs with gate metal work function either greater than the semiconductor work function, i.e. $w_M > w_S$, or smaller than the semiconductor work function, i.e. $w_M < w_S$. A low work function metal tend to induce electrons in the channel, tuning the channel to the charge accumulation regime; while a high work function metal can induce the channel into the charge depletion regime, all at zero gate bias. For the Al-gate and Pd-gate $MoS_2$ FETs reported in this work, the shift in threshold voltage is around 0.76 V, which changes the threshold voltage from negative to positive and confers $MoS_2$ both enhancement-mode and depletion-mode FETs (Fig. 2C and 2D in the manuscript).

The difference between the work functions of two metals on a dielectric is generally different from that in vacuum. This phenomenon may be characterized quantitatively by the *S* parameter, which accounts for dielectric screening. It can be calculated as the ratio between the effective metal work-function difference on a dielectric to that in vacuum:

$$\Phi_{M,eff} = \Phi_{CNL,d} + S\,(\Phi_{M,vac} - \Phi_{CNL,d})$$

where $\Phi_{M,eff}$ is the effective work function of the metal in a dielectric and $\Phi_{M,vac}$ is the work function of the same metal in vacuum. $\Phi_{CNL,d}$ is the charge neutrality level of the dielectric. The difference between the effective work functions of two metals can then be related to their difference in vacuum: $\Delta\Phi_{M,eff} = \Delta\Phi_{M,vac}$. Since $\Delta\Phi_{M,eff}$ and $\Delta\Phi_{M,vac}$ are about 0.76 and 1.04 eV, respectively, we have S ~ 0.7 for the metals on $HfO_2$, which agrees closely with the value reported in ref. S6.

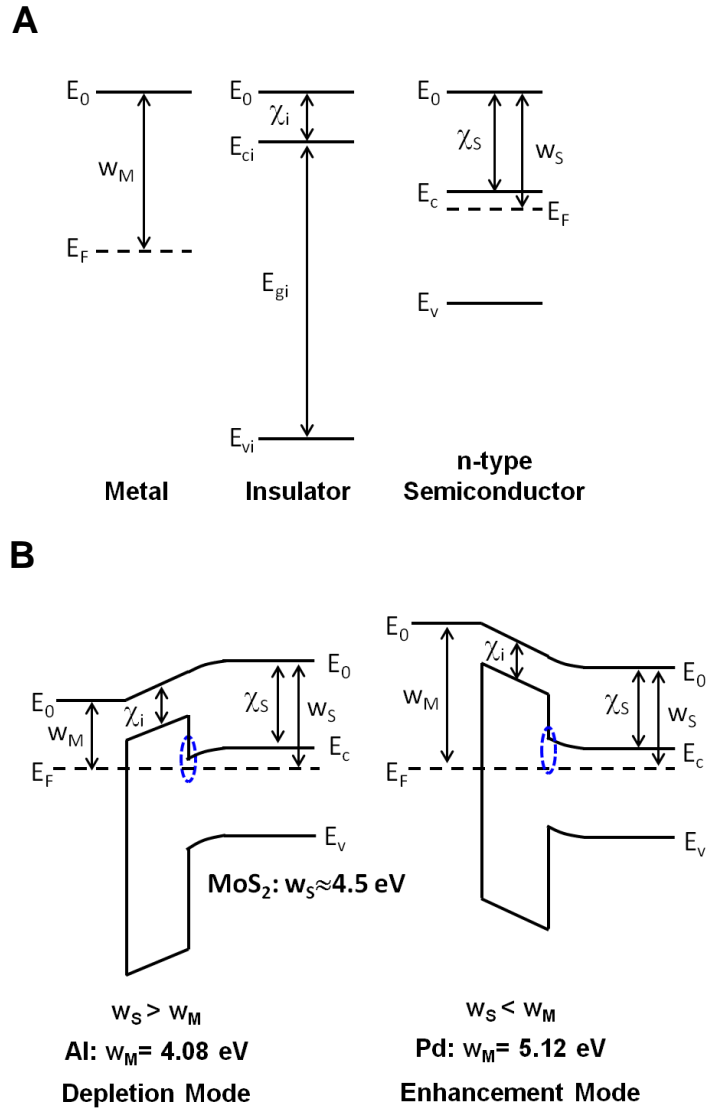

**Fig. S2.** Energy band diagrams **(A)** for isolated metal, insulator and semiconductor, and **(B)** after bringing them in intimate contact and thermal equilibrium is established. Depending on the different pairing of metal and semiconductor work functions, the metal-oxide-semiconductor (MOS) structure can induce the channel into either accumulation regime (for depletion mode FET) or depletion regime (for enhancement mode FET).

**Mobility Extraction**

The mobility of the bilayer thin films is extracted using the back-gate characteristics based on the expression $\mu = [dI_{ds}/dV_{bg}] \times [L/(WC_gV_{ds})]$. W=4 μm is the width of the device. L= 1 μm is the gate length. $C_g$ is the gate capacitance per unit area, which is based on 285 nm $SiO_2$ for the back gate. $V_{ds}$= 1V is the bias applied at the drain electrode relative to the source electrode.

Typical field effect mobility measured based on this two-contact method is around 5-15 $cm^2$ $V^{-1}$ $s^{-1}$ for bilayer $MoS_2$. After the deposition of $HfO_2$, the extracted mobility increases to above 300 $cm^2$ $V^{-1}$ $s^{-1}$ (Fig. S3). This significant increase in mobility after the deposition of high-k dielectric agrees with the observation in ref. S3, which is attributed to the suppression of Coulomb scattering by the high-k dielectric environment[S7] and spatial-confinement-induced modification of the acoustic phonon spectrum in bilayer $MoS_2$[S8]. The complete understanding of this improvement will, however, need further theoretical and experimental studies. The field-effect mobility extracted using this method represents a conservative estimate of the mobility value because of effect of contact resistance[S3,S5], which was not de-embedded in the measurements.

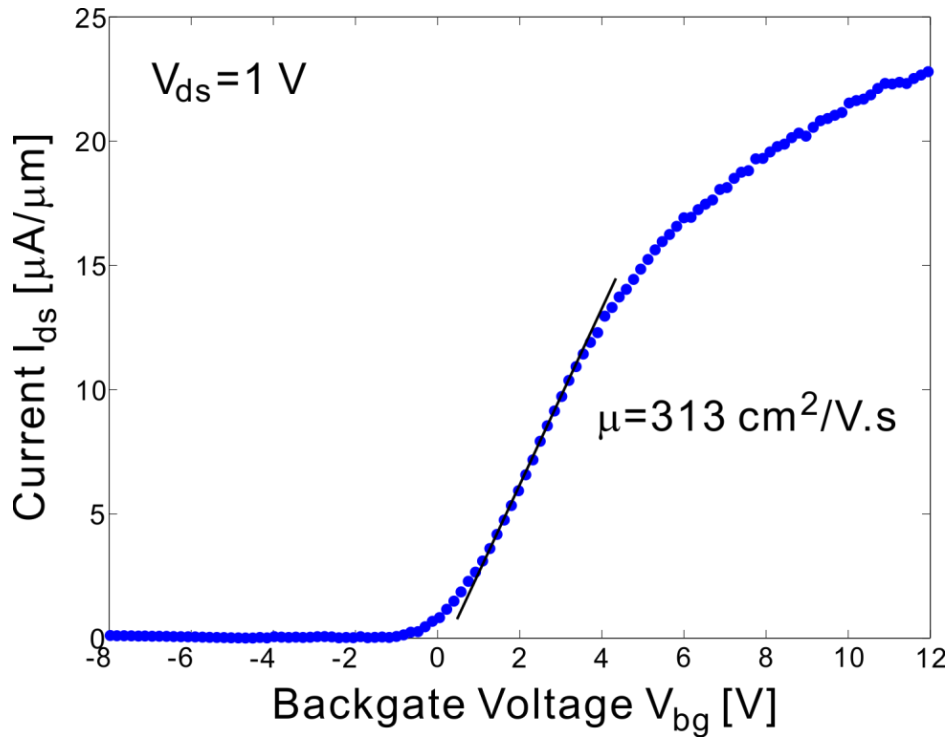

**Fig. S3.** Extraction of field-effect mobility based on back-gate characteristics of the device. The measurement is done at room temperature in vacuum (~$10^{-5}$ Torr).

**Estimation of the Ring Oscillator Speed**

We can estimate the expected oscillation frequency for the ring oscillator at $V_{dd}$=2 V based on the parasitic capacitances present at various parts of the circuit and the driving current supported by the FETs as follows[S9,S10]:

a. The average driving current $I_{ds}$ can be estimated from Fig. 3A of the report. Each inverter stage operates along the load line defined by the depletion-mode FET. Since the width of the FETs in the ring oscillator circuit is about 11 μm, we have $I \approx 22$ μA for $V_{dd}$=2 V.

The parasitic capacitance in the circuit is mainly contributed by two parts:

b. capacitances due to the overlap area between the top gate layers (M2 and M3) and the source/drain layer (M1). This capacitance is mainly due to the gate capacitance of the 12 FETs in the circuit as well as the overlap between the interconnects in the gate metal layers (M2 and M3) and that in the M1 layer. The gate capacitance of the 12 FETs are estimated based on their device width (about 11 μm on average for each FET), and their gate length (1μm for each FET) with 20 nm $HfO_2$ (dielectric constant ~22) as the dielectric material. This leads to capacitances of 1.285 pF. The remaining overlap area between the interconnects is around 5.3 μm$^2$, which gives an additional 0.051 pF. The total gate overlap capacitance is hence:

$$C_{ov\_gate} = 1.336 \text{ pF}$$

c. capacitances contributed by the conductive Si substrate with 285 nm $SiO_2$:

$$C_{ov\_Si} = 2.138 \text{ pF}$$

Hence, the estimated total parasitic capacitances per stage is equal to:

$$C \approx (C_{ov\_gate} + C_{ov\_Si})/6 = 0.579 \text{ pF}$$

The propagation delay per stage is estimated to be: $\tau_{pd} = CV_{dd}/I = 52.6$ ns.

Thus, at $V_{dd}$=2 V, the expected frequency for a 5-stage ring oscillator is equal to: $f = 1/(2n\tau_{pd}) = 1.9$ MHz.

This is very close to the measured value of the oscillation frequency, which is 1.6 MHz.